\documentclass[twocolumn,showpacs,prc]{revtex4}
\usepackage{graphicx}
\usepackage{dcolumn}
\usepackage{bm}
\begin{document}

\title{Electromagnetic Signals from Bacterial DNA}
\author{A. Widom and J. Swain}
\affiliation{Physics Department, Northeastern University, Boston MA USA}
\author{Y. N. Srivastava}
\affiliation{Physics Department \& INFN, University of Perugia, Perugia Italy}
\author{S. Sivasubramanian}
\affiliation{Nanoscale Technology and High Rate Manufacturing Research Center
\\ Northeastern University, Boston MA USA}

\begin{abstract}
Chemical reactions can be induced at a distance due to the propagation of electromagnetic 
signals during intermediate chemical stages. Although it is well known at optical frequencies, 
e.g. photosynthetic reactions, electromagnetic signals hold true for much lower frequencies as well. 
In E. coli bacteria such electromagnetic signals can be generated by electric transitions between 
energy levels describing electrons moving around DNA loops. The electromagnetic signals between 
different bacteria within a community is a ``wireless'' version of intercellular communication found 
in bacterial communities connected by ``nanowires''. The wireless broadcasts can in principle be of 
both the AM and FM variety due to the magnetic flux periodicity in electron energy spectra in bacterial 
DNA orbital motions.
\end{abstract}


\maketitle

\section{Introduction \label{intro}}

Biochemistry is most often described in terms of the short ranged molecular rearrangement interactions. 
However, it is clear that photo-induced biochemical reactions also exist. 
Photosynthesis constitutes an example of crucial biological importance. The photons which 
induce these chemical reactions can come from very distant sources (e.g. the sun). Chemical 
reactions can thereby be induced at a distance due to the propagation of electromagnetic 
signals during intermediate reaction stages. It appears reasonable to investigate the 
biochemical possibilities of electromagnetic signals at frequencies slow on the scale 
of light signals. Evidence for such reactions has been previously 
reported\cite{Benveniste:1996} wherein the time dependence of electromagnetic signals 
were recorded to later be employed at will.

In two recent and important experiments\cite{Mont1:2009,Mont2:2009}, it was shown 
that bacterial DNA macromolecules radiate electromagnetic signals which were  
monitored employing the voltage across an inductive pickup coil. The bacterial DNA 
within water was located in a test tube. The pickup coil was constructed with 
wires wrapped around the tube. 

Our purpose is to theoretically discuss the biophysical sources of these electromagnetic 
signals. The sources are argued to be due to electronic transitions between energy 
levels of electrons moving around the bacterial DNA loops.

One may deduce the spectral properties of the electromagnetic signals via electromagnetic  
noise; i.e. by employing the fluctuation dissipation theorem\cite{Landau:1980}
\begin{equation}
S_V(\omega )=\left(\frac{k_B T_{\rm noise}}{\pi}\right) {\Re e}Z(\omega +i0^+), 
\label{intro1}
\end{equation}  
wherein \begin{math} T_{\rm noise} \end{math} is the coil noise temperature and 
\begin{math} Z(\zeta ) \end{math} is the coil impedance as a function of complex 
frequency \begin{math} \zeta =\omega +i\sigma  \end{math}. Eq.(\ref{intro1}) is 
essential for the pickup coil method of detecting electromagnetic signals. 

\subsection{Low Frequency Noise \label{LFN}}

In the regime of very low frequencies, 
say in the range \begin{math} 1\ {\rm Hz}<(\omega /2\pi )< 20\ {\rm Hz} \end{math},
spectral noise can appear from the electronic magnetic moment precession due to 
small magnetic fields. Even if the shielding of magnetic fields due to external sources 
were perfect, the thermal fluctuations of coil currents would still give rise to 
magnetic fields yielding low frequency noise signals. The value of thermal magnetic fields 
\begin{math} B_T  \end{math} from thermal fluctuations is estimated in Sec.\ref{elf}.

\subsection{High Frequency Noise \label{HFN}}

In the regime of higher frequencies, say in the interval 
\begin{math} 0.2\ {\rm KHz} < (\omega /2\pi ) < 5\ {\rm KHz}  \end{math}, there exist 
sharp peaks in the noise spectral function \begin{math} S_V(\omega ) \end{math}. 
It is worthy of note that the bacterial DNA molecule is in the shape of a loop.
In what follows in Sec.\ref{lps}, we will model the coil noise spectra to the motion of nearly free 
electrons moving in a spatial loop of length \begin{math} L \end{math} which includes the helix 
DNA coils. The elementary Schr\"odinger equation will be solved in Sec.\ref{lps}. 

Fitting de Broglie electron waves on the loop, one finds the usual free electron quantum 
energy levels 
\begin{equation}
E_n=\left(\frac{2\pi^2\hbar^2}{mL^2}\right)n^2, 
\ \ \ \ \ \  n=0,\pm 1,\pm 2,  \cdots .
\label{intro2}
\end{equation}
The associated Bohr transition frequencies 
\begin{eqnarray}
\omega_n = \frac{E_{n+1}-E_{n}}{\hbar}=(2n+1)\varpi ,
\nonumber \\ 
\varpi =\left(\frac{2\pi^2\hbar }{mL^2}\right),
\label{intro3}
\end{eqnarray}
should then appear as broadcast electromagnetic signals at frequency increments
\begin{math} \Delta \omega /2\pi  = \varpi /\pi \end{math} 
in E. coli bacteria. This expectation is experimentally valid\cite{Mont1:2009,Mont2:2009}. The reader 
will note that these frequencies (not just the transition energies) depend on the 
(reduced) Planck's constant $\hbar$. Hence the derived result for the measured frequencies is indeed 
quantum mechanical, {\it not} classical. 

\subsection{Magnetic Moments \label{MM}}

The orbital magnetic moment of an electron moving around a baterial DNA loop may 
be understood as follows: (i) The mean velocity of electrons moving about the loop 
determines the electronic mean current according to 
\begin{eqnarray}
I_n=\frac{ev_n}{L}=\left(\frac{2\pi e\hbar }{mL^2}\right)n,
\nonumber \\ 
I_n=\frac{e}{T_n}=ef_n=e\left(\frac{\varpi }{\pi}\right)n, 
\nonumber \\ 
n=0,\pm 1,\pm 2,  \cdots  .
\label{intro4}
\end{eqnarray} 
wherein \begin{math} T_n =1/f_n \end{math} is the time period for the 
electron to go once around the ring.
(ii) The magnetic moment \begin{math} {\bf m} \end{math} of a circulating current 
loop around a vector area \begin{math} {\bf \Sigma} \end{math} is given by 
\begin{math} {\bf m}=I{\bf \Sigma}/c \end{math}; i.e. the mean magnetic moment is 
\begin{equation}
{\bf m}_n=\frac{I_n {\bf \Sigma }}{c}=
\left(\frac{\hbar e}{2mc}\right)\left[\frac{4\pi {\bf \Sigma }}{L^2}\right]n, 
\ \  n=0,\pm 1,\pm 2,  \cdots  .
\label{intro5}
\end{equation}  

If a magnetic field \begin{math} {\bf B} \end{math} is applied to the biological 
sample, then a magnetic flux 
\begin{equation}
\Phi ={\bf B\cdot \Sigma}
\label{intro7}
\end{equation}
will thread through the bacterial DNA loop. The orbital electronic energy levels 
will then exhibit a flux periodicity; i.e. with flux quantum period  
\begin{eqnarray}
\Phi_0=\frac{2\pi \hbar c}{|e|}\approx 4.13567\times 10^{-7}\ {\rm Gauss\ cm^2},
\nonumber \\ 
\Phi_0\approx 41.3567\ {\rm Gauss\ micron^2}. 
\label{intro8}
\end{eqnarray} 
The periodicity of the energy levels is thereby   
\begin{equation}
E_n(\Phi +\Phi_0)=E_n(\Phi ).
\label{intro9}
\end{equation}
Note that the magnetic flux quantum in Eq.(\ref{intro8}) is {\em twice} the value 
found for superconducting loops wherein the charge value is the electron pairing   
value \begin{math} q=2e \end{math}. 

As a consequence of Eq.(\ref{intro9}). magnetic field noise must be very small to 
observe the KHz signals. In general,  the magnetic flux threading the DNA loop,  
\begin{math} \Phi ={\bf B\cdot \Sigma } \end{math}, depends on how the vector 
area \begin{math} {\bf \Sigma } \end{math} is oriented with respect to the magnetic 
field \begin{math} {\bf B } \end{math}. The sharp spectral lines would thereby be 
considerably broadened. The broadening is discussed in Sec.\ref{FP}. 

\subsection{Bacterial Communication \label{BC}}
There has been considerable interest in bacterial communities wherein a bacterium 
is connected to neighboring bacteria by means of {\em narrow 
nanowires}\cite{Ntarlagiannis:2007,El-Naggar:2008,El-Naggar:2010}. It is believed 
that the purpose of the nanowires is to allow for intercellular electronic 
communications. More advanced on the evolutionary scale are the more modern bacterial  
communities which are {\em wireless}. The electromagnetic signals sent from a 
bacterium to neighboring bacteria can be due to relatively low frequency electron 
level transitions within the DNA.

\section{Magnetic Moment Precession \label{elf}}

If a compact object with magnetic moment \begin{math} {\bf m} \end{math} 
is subject to a magnetic field \begin{math} {\bf B} \end{math}, then a 
precession occurs, 
\begin{equation}
\frac{d {\bf m}}{dt}=\gamma {\bf m}\times {\bf B},
\label{elf1}
\end{equation} 
at a rotational angular velocity  
\begin{equation}
2\pi {\bf f}\equiv {\bf \Omega}=-\gamma {\bf B},
\label{elf2}
\end{equation} 
wherein \begin{math} \gamma \end{math} is the gyromagnetic ratio. The 
gyromagnetic ratio associated with electronic charged current  
flows is given by  
\begin{equation}
\gamma =\frac{e}{2mc}\approx -\frac{(8.7941\times 10^6)}{\rm Gauss\ sec}.
\label{elf3}
\end{equation} 
For a cylindrical water sample of volume \begin{math} V  \end{math} within an 
inductive pickup coil, one may employ the magnetic field equipartition 
theorem\cite{Landau:1980} in the form 
\begin{eqnarray}
\frac{V\overline{B^2}}{8\pi }=\frac{k_BT}{2}\ ,
\nonumber \\ 
B_T=\sqrt{\frac{4\pi k_BT}{V}}\ .
\label{elf4}
\end{eqnarray}
From Eqs.(\ref{elf2}) and (\ref{elf4}), one obtains a magnetic frequency,  
\begin{math} 2\pi f_T=|\gamma |B_T  \end{math}, which serves as a lower bound for 
the magnetic noise frequency within the coil. Numerically,   
\begin{eqnarray}
f_T\approx 1.01\ {\rm Hz}\ 
\sqrt{\left(\frac{\rm cm^3}{V}\right)\left(\frac{T}{300\ {\rm ^oK}}\right)}
\ \ \ ({\rm orbital}),
\nonumber \\ 
f^{spin}_T\approx 2.02\ {\rm Hz}\ 
\sqrt{\left(\frac{\rm cm^3}{V}\right)\left(\frac{T}{300\ {\rm ^oK}}\right)}
\ \ \ ({\rm spin}),
\label{elf5}
\end{eqnarray} 
wherein the electron spin gyromagnetic ratio involves the factor \begin{math} g\approx 2 \end{math}.
The above are consistent with the experimental value\cite{Mont1:2009,Mont2:2009} of 
\begin{math} \sim 7\ {\rm Hz} \end{math} for diluted bacterial DNA samples in water.
The magnetic field value in Eq.(\ref{elf4}) is internal to the system and will hold true 
if {\em external} magnetic fields are sufficiently well {\em screened}.

\section{Electrons and DNA Loops  \label{lps}}

Consider an electron which can move around a bacterial DNA loop including the windings of helices 
and along the ordered water layers within which the DNA is stored. A reasonable Hamiltonian  
model for the electronic energy levels may then be written as 
\begin{eqnarray}
V(s+L)=V(s)\ \ \ {\rm and}\ \ \ A(s+L)=A(s),
\nonumber \\ 
H=\frac{1}{2m}\left(-i\hbar \frac{d}{ds}-\frac{e}{c}A(s)\right)^2+V(s),
\label{lps1}
\end{eqnarray} 
wherein allowed electron wave functions have a periodicity around the DNA loop 
\begin{equation}
\psi (s+L)=\psi (s). 
\label{lps2}
\end{equation}
The vector potential tangent to the closed loop may be written as the periodic function 
\begin{equation}
A(s)=\frac{\Phi }{L}+\sum_{n=1}^\infty A_n\cos\left(\frac{2\pi n s}{L}+\theta_n\right)
\label{lps3}
\end{equation}
In virtue of the gauge transformation 
\begin{eqnarray}
\chi (s)=-\sum_{n=1}^\infty \left(\frac{LA_n}{2\pi n}\right)\sin\left(\frac{2\pi n s}{L}+\theta_n\right), 
\nonumber \\ 
A(s)\to A(s)+\frac{d\chi (s)}{ds}\ ,
\nonumber \\ 
\psi (s) \to e^{(ie\chi(s) /\hbar c)}\psi(s),  
\label{lps4}
\end{eqnarray}
The Hamiltonian in Eq.(\ref{lps1}) may be written with boundary conditions as 
\begin{eqnarray}
H=\frac{1}{2m}\left(-i\hbar \frac{d}{ds}-\frac{e\Phi}{cL}\right)^2+V(s), 
\nonumber \\ 
V(s+L)=V(s), 
\nonumber \\ 
\psi (s+L)=\psi(s).
\label{lps5}
\end{eqnarray}

For harmless E. coli K-12 bacteria, the loop length is \begin{math} 4,639,221\ bp  \end{math}, or in 
absolute length units  
\begin{equation}
\tilde{L}=0.157733514\ {\rm cm}.   
\label{lps6}
\end{equation} 
If the mobile electron moving around the DNA in the ordered water layer, skips rungs around the helices, 
then the electron path around the DNA would be considerably shorter than in Eq.(\ref{lps6}) We find 
satisfactory agreement between the electron spectra and the observed pickup coil noise with the 
shorter length scale 
\begin{equation}
L\approx 0.086\ {\rm cm},   
\label{lps7}
\end{equation} 
yielding for zero magnetic flux and for uniform potential 
\begin{equation}
\Delta f=\frac{2\pi \hbar}{mL^2}=\frac{\varpi }{\pi}\approx 1\ KHz .
\label{lps8}
\end{equation} 
In order to resolve the energy spectra, the magnetic field must be carefully screened.

Modes where the electron skips rungs and is delocalized over a longer distance may
also be present, possibly with a strong dependence on hydration and the water
shell around DNA in physiological systems, but would correspond
to much higher frequencies. It would be interesting to search for these also.

\section{Flux Periodicity \label{FP}}

For the general spectral model in Eq.(\ref{lps5}), the energy levels give rise to a magnetic flux periodicity 
as in Eq.(\ref{intro9}). Different bacteria will exhibit different spectra depending on the orientation of the 
DNA loop vector area \begin{math} {\bf \Sigma}  \end{math} with respect to the magnetic field 
\begin{math} {\bf B}  \end{math}. 
From the flux periodicity and Faraday's law, 
\begin{equation}
E(\Phi +\Phi_0)=E(\Phi )\ \ \ {\rm and}\ \ \ V=-\frac{1}{c}\left(\frac{d\Phi}{dt}\right),
\label{FP1}
\end{equation}
one finds voltage modulated sidebands at the electronic spectral frequencies determined by the Faraday 
law voltage around the DNA loop via the resonance condition\cite{Giudice:1989} 
\begin{equation}
\omega_V=\frac{eV}{\hbar}\ .
\label{FP2}
\end{equation}
The wireless communications can thereby be transmitted for both AM and FM broadcast systems.

\section{Experimental Tests\label{ET}}

In addition to explaining the results of some already-performed experiments, we note
that the work here also makes concrete predictions which we hope will be tested in the
laboratory. Transition frequencies scale as the square of the reciprocal of the number of
base pairs in the DNA (see equation 3), so the DNA of different organisms should exhibit
different frequency spectra in a fashion which is directly comparable. Also of interest is
the DNA in plasmids, which is much shorter and should allow a test of the validity of equation
3 with diverse length scales. The very low frequencies attributed to magnetic moment precession
scale as the square root of temperature divided by volume and both parameters should be readily
variable even in the present experimental setups. In addition, the background RMS magnetic
field ($\bar{B^2}$) in equation 12 could be supplemented by a larger external field and the
corresponding (higher) frequencies determined from $2\pi f_T =|\gamma| B_T$ searched for.

If the predictions in this paper are borne out, it may even be possible to determine the
length of the DNA present in a sample directly from measurements of the corresponding
frequencies. While the length of the DNA of an organism cannot be expected to uniquely
define it, such a non-invasive partial determination of DNA structure may have implications
for biological and medical tests.

The shorter wavelengths associated with possible delocalized electron modes that ``skip
rungs'' mentioned at the end of section \ref{lps} would also be interesting to look for.

Finally we note that it may be interesting not merely to listen in on DNA, but to use
external time-varying fields to excite it at its resonant frequencies. What effects such
excitation might have on living cells is far from obvious, but could indicate specific
mechanisms for non-ionizing electromagnetic radiation with quanta of very low energy
(even below $kT$ at room temperature) to have biological effects.

\section{Conclusions}

Although biochemical reactions are often described in terms of molecular contacts, electromagnetic signals 
can often be employed to allow chemical reaction control at a distance. The photosynthetic reactions are a 
classic case of chemical reaction control via electromagnetic signal propagation. At frequencies much less than 
the optical, there is a clear electromagnetic signal propagation in E. coli bacterial communities. We have probed a 
model wherein such signals are due to quantum electronic transitions of electrons in orbital motion about DNA loops.

The electromagnetic signals between different bacteria within a 
community is a ``wireless'' version of intercellular communication found in bacterial communities 
connected by ``nanowires''. The wireless broadcasts can in principle be of both the AM and FM variety 
due to the magnetic flux periodicity in electron energy spectra in bacterial DNA orbital motions.
AM signals can arise from the Bohr transition frequencies between different electronic energy orbitals about 
the DNA loops. FM modulation signals can arise from the Faraday law voltage controlled signal modulation 
frequency in Eq.(\ref{FP2}). There is considerable work required to extract the bioinformation contained in these 
electromagnetic signals.

\section{Acknowledgments\label{Ack}}

YS would like to thank Professor Luc Montagnier for many valuable discussions regarding the data from his group.


\begin{thebibliography}{11}


\bibitem{Benveniste:1996}
J. Benveniste, P. Jurgens, J. A\"issa, 
{\it Faseb Journal} {\bf 10 A}, 1479 (1996).

\bibitem{Mont1:2009}
L. Montagnier, J. A\"issa, S. Ferris, J-L. Montagnier and C. Lavallee,
{\it Interdiscip. Sci. Comput. Life Sci.} {\bf 1}, 81 (2009).

\bibitem{Mont2:2009}
L. Montagnier, J. A\"issa, C. Lavallee, M. Mbamy, J. Varon and H. Chenal,
{\it Interdiscip. Sci. Comput. Life Sci.} {\bf 1}, 245 (2009).

\bibitem{Landau:1980}
L.D. Landau and E.M. Lifshitz, ``Statistical Physics'', Part I, 3rd Edition, 
Chapter XII, Pergamon Press, Oxford (1980).

\bibitem{Ntarlagiannis:2007}
D. Ntarlagiannis, E.A. Atekwana, 3. Eric, A. Hill, and Yuri Gorby, 
{\it GeoPhys. Res. Lett.} {\bf 34}, L17305 (2007).

\bibitem{El-Naggar:2008}
M.Y. El-Naggar, Y.A. Gorby, W. Xia and K.H. Nealson, 
{\it Biophys. J. : Biophys. Letts.} L10 (2008). 

\bibitem{El-Naggar:2010}
M.Y. El-Naggar, G. Wangerb, K.M. Leungc, T.D. Yuzvinskya, G. Southame, J. Yangc,
W.M. Laud, K.H. Nealsonb and Y.A. Gorbyb
{\it PNAS} {\bf 107}, 18127 (2010).

\bibitem{Giudice:1989}
E. Del Guidice, S. Doglia, M. Milani, C.W. Smith and G. Vitiello, 
{\it Physica Scripta} {\bf 40}, 786 (1989).






\end{thebibliography}
\end{document}